\documentclass[prb,twocolumn,showpacs,preprintnumbers,amsmath,amssymb]{revtex4}

\usepackage{color}
\usepackage{graphicx}
\usepackage{dcolumn}
\usepackage{bm}

\preprint{submitted to Journal of Applied Physics}

\begin{document}

\title{Robustness and stability of half-metallic ferromagnetism in alkaline-earth metal
mononitrides against doping and deformation}

\author{K. \"Ozdo\u{g}an}\email{kozdogan@yildiz.edu.tr}
 \affiliation{Department of Physics, Yildiz Technical University, 34210
\.{I}stanbul, Turkey}

\author{E. \c{S}a\c{s}{\i}o\u{g}lu}\email{e.sasioglu@fz-juelich.de}

\affiliation{Peter Gr\"{u}nberg Institut and Institute for
Advanced Simulation, Forschungszentrum J\"{u}lich and JARA,
52425 J\"{u}lich, Germany \\
and Department of Physics, Fatih University, 34500,
B\"{u}y\"{u}k\c{c}ekmece, \.{I}stanbul, Turkey}

\author{I. Galanakis}\email{galanakis@upatras.gr}
\affiliation{Department of Materials Science, School of Natural
Sciences, University of Patras,  GR-26504 Patra, Greece}

\begin{abstract}
We employ ab-initio electronic structure calculations and study
the magnetic properties of CaN and SrN compounds crystallizing in
the rocksalt structure. These alkaline-earth metal mononitrides
are found to be half-metallic with a total spin magnetic moment
per formula unit of 1.0 $\mu_B$. The Curie temperature is
estimated to be 480 K for CaN and 415 K for SrN well-above the
room temperature. Upon small degrees of doping with holes or
electrons, the rigid-band model suggests that the magnetic
properties are little affected. Finally we studied for these
alloys the effect of deformation taking into account
tetragonalization keeping constant the unit cell volume which
models the growth on various substrates. Even large degrees of
deformation only marginally affect the electronic and magnetic
properties of CaN and SrN in the rocksalt structure. Finally, we
show that this stands also for the zincblende structure. Our
results suggest that alkaline-earth metal mononitrides are
promising materials for magnetoelectronic applications.
\end{abstract}


\maketitle

\section{Introduction}\label{sec1}

Half-metallic ferromagnetic or ferrimagnetic materials have been
extensively studied during the last decade due to their potential
application in magnetoelectronic devices.\cite{Zutic,Felser,Zabel}
These magnetic systems present a usual metallic behavior
concerning their majority-spin electrons and a semiconducting gap
in the minority-spin electronic band structure and thus 100\%\
spin-polarization at the Fermi level and could in principle
maximize the efficiency of spintronic devices.\cite{Katsnelson}.
As expected the main interest has been focused in the potential
half-metallic magnets based on transition-metal elements but the
latter due to their large spin-magnetic moments are expected to
present large external stray magnetic fields and thus devices
based on them should exhibit considerable energy losses. A way to
detour this problem is to search for new materials presenting much
smaller spin magnetic moments and to this respect a very
interesting case are ferromagnetic compounds which do not contain
transition-metal atoms. These are widely known in literature with
various names like $d^0$ magnets, $p$-ferromagnets or sp-electron
ferromagnets.\cite{Volniaska10} Several of them combine
ferromagnetism with the desired for applications half-metallicity.

There are several ways to create sp-electron ferromagnets and an
extensive review is given in Ref. \onlinecite{Volniaska10}. First,
we can induce vacancies or holes at the cation sites in oxides,
which  lead to a small exchange-splitting of the cation spin-up
and spin-down states and thus to half-metallic
ferromagnetism.\cite{Volniaska10,Zhou12,Uchino12,Maca08,Zheng11,Guan10}
A second route to $p$-magnets are the so--called molecular
solids.\cite{Volniaska10} In these materials the oxygen or
nitrogen atoms form dimers which are ferromagnetically or
antiferromagnetically coupled between
them.\cite{Ylvisaker10,Kovacik09,Kim10,Volniaska08,Slipukhina11,Wu10}
The third way to create half-metallic sp-electron ferromagnets is
the doping of oxides with nitrogen atoms.\cite{Volniaska10} N
impurities at anionic sites (oxygens) present a splitting of their
$p$-bands and the majority-spin states are completely occupied
while the minority-spin states are partially occupied leading to
half-metallicity.\cite{Mavropoulos09,Xiao10,Banikov11,Yang12,Kenmonchi04,Kenmonchi04b}
Experimental evidence for the occurrence of magnetism in N-doped
MgO has been provided by Liu and collaborators.\cite{Liu11}

An alternative way to half-metallic $sp$-electron ferromagnets is
the growth of I/II-IV/V nanostructures in metastable lattice
structures similar to the case of transition-metal pnictides and
chalcogenides in the metastable zincblende
structure.\cite{ReviewCrAs} Several studies to this research
direction have appeared following the pioneering papers published
by Geshi et al\cite{Geshi04} and Kusakabe et al\cite{Kusakabe04}
who have shown using first-principles calculations that CaP, CaAs
and CaSb alloys present half-metallic ferromagnetism when grown in
the zincblende structure. Ca atom provides two valence electrons
(occupying the 4s states in the free atom) while the anions
(P,As,Sb) provide 5 valence electrons (\textit{e.g.} in free atom
of As the atomic configuration is 4s$^2$ 4p$^3$). In total there
are 7 valence electrons per unit cell. The first two occupy the
s-valence states created by As atoms which lie deep in energy. The
p-states of anions hybridize strongly with the triple-degenerated
t$_{2g}$ d-states of Ca, which transform following the same
symmetry operations, and form bonding and antibonding hybrids
which are separated by large energy gaps. The bonding hybrids
contain mostly p-admixture while the antibonding hybrids are
mainly of d-character. The remaining 5 valence electron occupy the
bonding hybrids which are mainly of anionic p-character in such a
way that all three majority-spin p-states are occupied while in
the minority-spin band the Fermi level cross the bands so that
only the two out of three p-states are occupied. This gives in
total a spin magnetic moment per formula unit of exactly 1
$\mu_B$. This mechanism is similar for all half-metallic
ferromagnetic I/II-IV/V alloys in all three zincblende (ZB),
wurtzite (WZ) and rocksalt (RS) metastable structures, and the
spin magnetic moment follows a Slater-Pauling behavior with the
total spin magnetic per formula unit in $\mu_B$ being 8 minus the
number of valence electron in the unit cell: $M_t=8-Z_t$. Evidence
of the growth of such nanosctructures has been provided by Liu et
al who have reported successful self-assembly growth of ultrathin
CaN in the rocksalt structure on top of Cu(001).\cite{Liu08}
Finally we have to note that materials containing C or N seem to
be more promising for applications since the Hund energy for the
light atoms in the second row of the periodic table is similar to
the Hund energy of the 3d transition metal atoms.

Following the Refs. \onlinecite{Geshi04} and
\onlinecite{Kusakabe04} mentioned in the previous paragraph  a lot
of  studies on such compounds have appeared based on first
principles calculations and we will give a short overview of them.
Although extensive studies exist also for the alkali metal
alloys\cite{Sieberer06,Zhang08,Zberecki09,Yan12,Gao09,Gao11} and
the alkaline earth
chalcogenides,\cite{Gao07,Gao07c,Gao07b,Dong11,Zhang08b,Gao09b,Verma10}
 most of the attention has been focused on the alkaline-earth metal
(IInd column) compounds with the Vth-column elements and mainly
the
nitrides.\cite{Geshi04,Kusakabe04,Sieberer06,Yao06,Li08,Volniaska07,Geshi07,Gao08,Droghetti09,Laref11,Gao11b}
Sieberer et al studied all possible II-V combinations in the ZB
and WZ structures and found that all alloys containing Ca, Sr and
Ba are half-metallic while only MgN was half-metallic between the
Mg-based compounds.\cite{Sieberer06} It was also shown in Ref.
\onlinecite{Sieberer06} that the ferromagnetic state is
energetically preferable to both the non-magnetic and the
antiferromagnetic configurations. Volnianska and Boguslawski, as
well as Geshi and collaborators have studied the alkaline-earth
metal nitrides and have shown that the RS is the more stable
structure with formation anergies of about -11 eV per unit
cell.\cite{Volniaska07,Geshi07} Gao et al have shown that among
the RS alloys containing Ca, Sr or Ba as a cation and N, P or As
as an anion only the nitrides are stable half-metallic
ferromagnets with a total spin magnetic moment of 1 $\mu_B$ and
cohesive energies about -9 eV per formula unit.\cite{Gao08}
Droghetti and collaborators have shown that RS-MgN is in verge of
the half-metallicity and suggested that MgN inclusion upon the
N-doping of MgO should lead to a material suitable for magnetic
tunnel junctions.\cite{Droghetti09} The most recent studies on
nitrides concern the Curie temperature in the ZB-structure which
was found to be 430 K in CaN,\cite{Laref11} and the RS-CaN/ZB-InN
and RS-SrN/ZB-GaP (111) interfaces which were found to retain
half-metallicity only when the interface is made up from Ca-N or
N-In atoms in the first case and N-Ga in the second
case.\cite{Gao11b}

\section{Motivation and computation method}\label{sec2}

As we can conclude from the discussion in the previous section,
rocksalt alkaline-earth metal nitrides combine some unique
properties among these alloys : they have a small spin magnetic
moment per formula unit (1 $\mu_B$) and thus create small external
magnetic fields, (ii) they present very stable half-metallicity
upon hydrostatic pressure, (iii) their equilibrium lattice
constant are close to a lot of semiconductors, (iv) results on the
ZB-structure suggest high values of the Curie temperature also for
the RS-structure, (v) the half-metallic gaps are wide, and (vi)
interfaces with semiconductors retain half-metallicity. Thus in
this manuscript we complete the previous studies on the RS CaN and
SrN alloys focusing on properties which have not yet been
determined. In the first part of our study we employ the augmented
spherical waves method (ASW)\cite{asw} within the atomic--sphere
approximation (ASA)\cite{asa} in conjunction to the generalized
gradient approximation (GGA) for the exchange-correlation
potential \cite{gga} to perform ground state electronic structure
calculations for RS CaN and SrN using the lattice constants from
Ref. \onlinecite{Geshi07} (5.02 \AA\ for CaN and 5.37 \AA\ for
SrN). Notice that within ASW, empty spheres have been used, where
needed, in order to describe better the lattice filling. We use
these results and the frozen-magnon technique\cite{magnon} to
determine the exchange constants and Curie temperature in both the
mean-field (MFA) and random-phase (RPA) approximations. The
formalism has been already presented for a one-sublattice system
like the ones under study (only N-N interactions contribute since
the Ca(Sr) atoms have negligible spin magnetic moments) in Ref.
\onlinecite{Sasioglu}. Then using a rigid band model as in Ref.
\onlinecite{Galanakis} we study how the exchange constants, Curie
temperature, spin-polarization and spin magnetic moments vary with
the band-filling. In the second part of our study we employ the
full--potential nonorthogonal local--orbital minimum--basis band
structure scheme (FPLO)\cite{koepernik} within the GGA
approximation\cite{gga} to determine the equilibrium lattice
constants of rocksalt CaN and SrN which are found to be almost
identical to the results of Geshi et al.\cite{Geshi07} We,
thereafter, study the effect on the electronic and magnetic
properties of  tetragonalization keeping the volume of the unit
cell constant which usually accounts for growth on different
substrates. To make this study more complete we compare these
results with the results also on the zincblende lattice structure.

\begin{table}
\caption{Calculated magnetic properties for the CaN and SrN alloys
in the rocksalt structure using the ASW method. The spin magnetic
moments are given in $\mu_B$ units and the Curie temperatures in
Kelvin. The lattice constants are the ones determined in Ref.
\onlinecite{Geshi07}.}
\begin{ruledtabular}
\begin{tabular}{llllllccc}
& a(\AA )& $m^\mathrm{Ca(Sr)}$ & $m^\mathrm{N}$ &
$m^\mathrm{Total}$ & T$_\textmd{C}^{\textmd{MFA}}$
&T$_\textmd{C}^{\textmd{RPA}}$

\\ \hline
CaN & 5.02 &     0.015   &  0.980                    &  1.00           &  620    & 480 \\
SrN & 5.37 &     0.007   &  0.995                  &  1.00           &  594     & 418  \\
\end{tabular}
\end{ruledtabular}
\label{table1}
\end{table}

\section{Robustness of Curie temperature against
doping}\label{sec3}

We will start the presentation of our results from the temperature
related properties. As discussed above we employed the ASW method
to study the electronic properties of both CaN and SrN in the
rocksalt structure using as lattice parameters the equilibrium
ones from Ref. \onlinecite{Geshi07}. In Table \ref{table1} we have
gathered the calculated magnetic properties. The spin magnetic
moments present similar behavior as in the studies discussed in
Section \ref{sec1}. Both compounds present a total spin magnetic
moment per formula unit of 1 $\mu_B$ in agreement with the
Slater-Pauling rule for half-metallic mononitrides (there are in
total 7 valence electron per unit cell) and thus we expect the
density of state (DOS) to present half-metallic properties. Almost
all the moment is concentrated at the N atoms. Ca atom carries a
spin magnetic moment of only 0.015 $\mu_B$ and Sr of 0.007
$\mu_B$. Thus the cation-cation as well as the cation-nitrogen
interactions should make a minimal contribution to the exchange
constants and it is enough to consider the N-N intrasublattice
magnetic interaction when discussing the temperature related
properties. The very small spin magnetic moments at the cationic
site reflect the very small charge at these sites since the
occupied bonding hybrids are mainly of anionic character having
only a very small cationic d-admixture as discussed in Section
\ref{sec1}.

\begin{figure}
\centering
\includegraphics[width=\columnwidth]{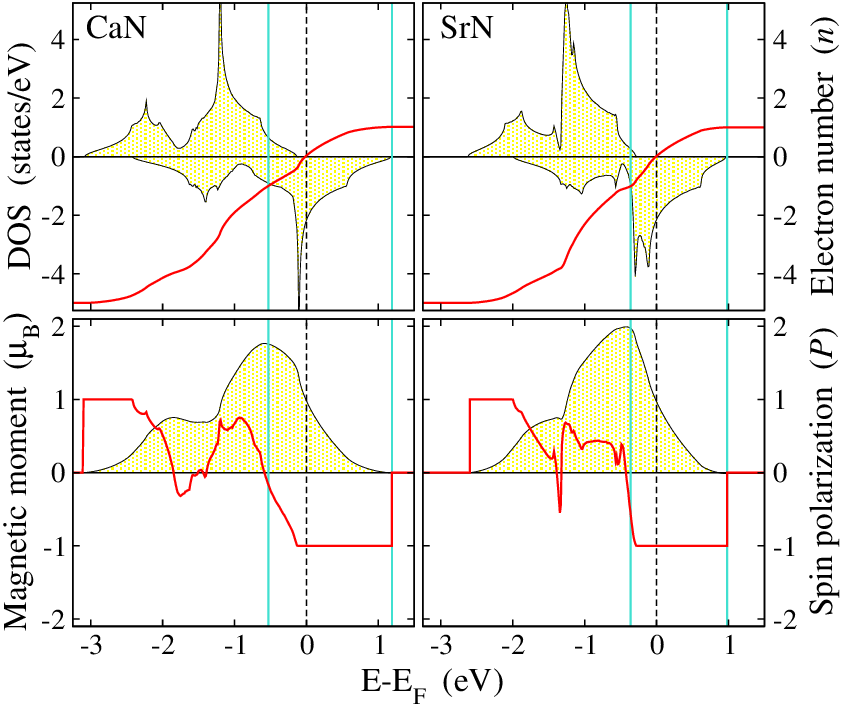}
\caption{(Color online) Upper panel: Total DOS (shaded region) for
both CaN and SrN alloys in the rocksalt structure using the ASW
method. The red line represents the electron counting setting as
zero the electrons at the Fermi level. With cyan vertical
solid lines we denote the cases of $\pm$1 electron.\\
Lower panel: total magnetic moment (shaded region) and
spin-polarization (red line) as a function of the electron
counting.} \label{fig1}
\end{figure}

The total DOS for both compounds are presented in Fig. \ref{fig1}.
We do not present the deep-lying occupied s-states since they are
located at about -12 eV below the Fermi level and thus are not
relevant for the discussion of the electronic properties. The
bonding and the antibonding hybrids are separated by a gap which
is about 1.7 eV for the minority-spin states; the antibonding
hybrids are not presented in the figure. Note also that the
antibonding states present almost no exchange splitting.
Ca(Sr)-resolved DOS is very small with respect to the N-resolved
DOS in the energy window, of the occupied states and thus we can
safely assume that the total DOS presented in the figure coincides
in this energy window with the N-resolved DOS. The Fermi level
cross the minority-spin DOS while all the majority N p-states are
occupied and the Fermi level falls within a majority-spin gap
which is 3.1 eV wide for CaN and 3.0 for SrN. The so-called
half-metallic gap, defined as the energy distance between the
highest occupied majority-spin state and the Fermi level, is about
0.1 eV for CaN and 0.2 eV for SrN.

Small degrees of doping can be assumed to result in small shifts
of the Fermi level as assumed also for the semi-Heusler compounds
in Ref. \onlinecite{Galanakis}. In Fig. \ref{fig1} we present in
the upper panel the total DOS for both CaN and SrN alloys and with
the red line the electron counting setting the number of electrons
at the Fermi level as zero. The vertical blue lines denote the
limits of $\pm$1 electron. As we dope our system with electrons
and we move to higher values of the energy with respect to the
Fermi level we populate also the minority-spin bonding hybrids.
For exactly a surplus of one electron per formula unit all
minority-spin bonding hybrids are occupied (we remind here that in
these alloys two out of three minority-spin p-states were already
occupied) and we end up with a compound with zero total spin
magnetic moment This situation is probably unphysical since it
corresponds to a very large degree of doping and ScN and YN in the
rocksalt structure which have one more valence electron than CaN
and SrN are semiconductors. If we start doping CaN and SrN with
holes we move deeper in energy. The two alloys present a
significant difference in their behavior due to the larger
exchange-splitting in the case of SrN which is also reflected on
the larger half-metallic gap. As shown by the spin-polarization
presented with the red line in the lower panel of Fig. \ref{fig1},
for CaN small doping with holes results very quickly in loss of
half-metallicity and the spin polarization deviates from the
perfect 100\%. On the other hand in SrN even doping with one hole
preserves the half-metallic character and the perfect
spin-polarization since the Fermi level is at the verge of the
occupied majority-spin electronic bands. In the lower panel of the
same figure we also present the variation of the spin-magnetic
moment with the electron counting which also reflects our
discussion. Exactly at the Fermi level we have a total spin
magnetic moment of 1 $\mu_B$ and when we dope with holes it
increases. For one hole (corresponding to -1 electron in the
counting) the spin  magnetic moment of CaN reaches a value of 1.76
$\mu_B$ while for SrN the hole populates only minority-spin states
and the spin moment reaches the value of 2 $\mu_B$ expected from
the Slater-Pauling rule for perfect half-metals. Of course doping
with electrons leads to a decrease of the total spin magnetic
moment which vanishes when we add exactly one electron to our
system and all bonding hybrids are occupied.

\begin{figure}
\centering
\includegraphics[width=\columnwidth]{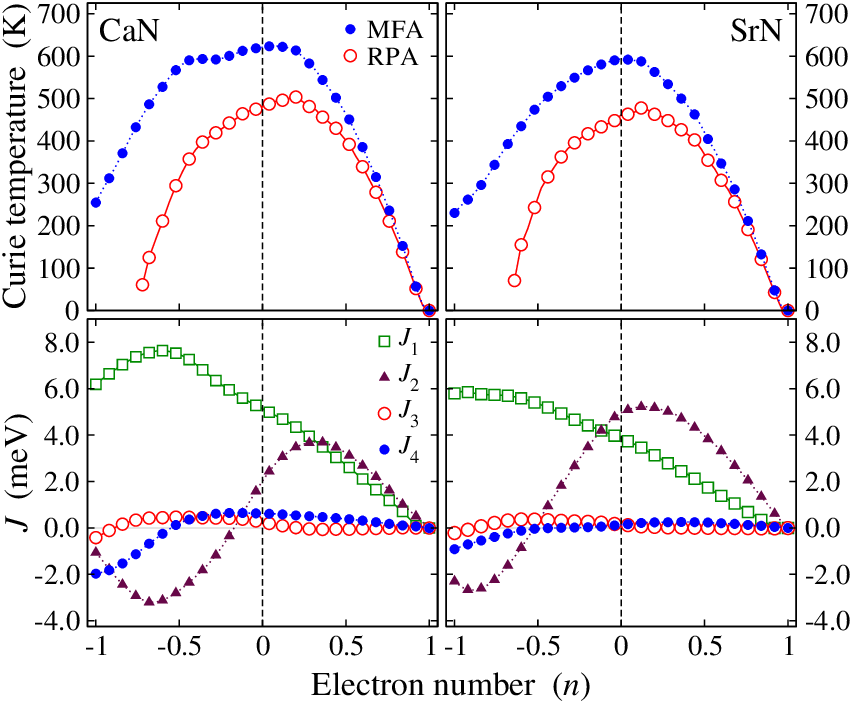}
\caption{(Color online) Upper panel: Calculated Curie temperatures
within both the random-field (RPA) and mean-field (MFA)
approximations for CaN and SrN in the rocksalt structures as a
function of the electron counting (see Fig. \ref{fig2}).\\
Lower panel: Calculated exchange constants between nitrogen atoms
as a function of the electron counting up to fourth neighbors. The
N-Ca(Sr) and Ca(Sr)-Ca(Sr) exchange constants are negligible with
respect to the N-N ones due to the very small spin magnetic moment
of the Ca(Sr) atoms.} \label{fig2}
\end{figure}

In the last part of this section we will concentrate on the Curie
temperatures. In Table \ref{table1} we present the estimated
values of the Curie temperature in Kelvin. The mean filed
approximation (MFA) gave a value of 620 K for CaN and 594 K for
SrN while the random-phase approximation (RPA) gave values of 480
K and 418 for CaN and SrN, respectively. RPA is expected to give
more accurate results with respect to MFA since RPA corresponds to
a larger weight of the lower-energy excitations contrary to MFA
which assumes an equal weight for both low- and high-energy
excitations.\cite{Sasioglu} The calculated Curie temperatures
exceed significantly the room temperature as was also the case for
various sp-electron ferromagnets discussed in Section \ref{sec1}
and thus these alloys can have potential room-temperature
applications in spintronic devices. Note that for CaN in the
zincblende structure we had calculated within RPA in Ref.
\onlinecite{Laref11} a value for the Curie temperature of 415 K
which is lower than the 480 K for the rocksalt-CaN in the present
study although the nitrogen atoms have a spin magnetic moment of
0.98 $\mu_B$ in both ZB and RS structures.

In Fig. \ref{fig2} we present how the calculated Curie temperature
and the exchange constants in the N-sublattice behave with respect
to the electron counting. We focus in a window of $\pm$1 electron
although usual doping with electrons or holes should result to a
much smaller change in the electronic counting. In all cases MFA
results are higher than the RPA ones and as we move to the case +1
electron the MFA and RPA values coincide. At the vicinity of the
zero electron counting the estimated Curie temperature still
exceeds considerably the room temperature For both CaN and SrN the
calculated RPA temperature reaches the room temperature either for
doping with 0.6 electrons or for doping with 0.5 holes and thus
for moderate degrees of doping we are well above the room
temperature. At the zero of the electron counting we get the
maximum Curie temperature due to the combination of the
contribution of the exchange constants between both N-N nearest
($J_1$) and next-nearest ($J_2$) neighbors as shown in the lower
panel of Fig. \ref{fig2}. Exactly at zero electron counting both
$J_1$ and $J_2$ favor ferromagnetism and make important
contribution to the Curie temperature. In SrN with respect to CaN
the $J_2$ takes larger values while the opposite occurs for $J_1$.
As we dope with electrons $J_1$ starts dropping fast, while when
we dope with holes $J_1$ increases considerably but at the same
time $J_2$ starts favoring antiferromagnetism leading also to
smaller estimated Curie temperatures. The interaction between N-N
third and fourth neighbors are significant only at the vicinity of
the -1 electron value for the electron counting.

\begin{figure}
\includegraphics[width=\columnwidth]{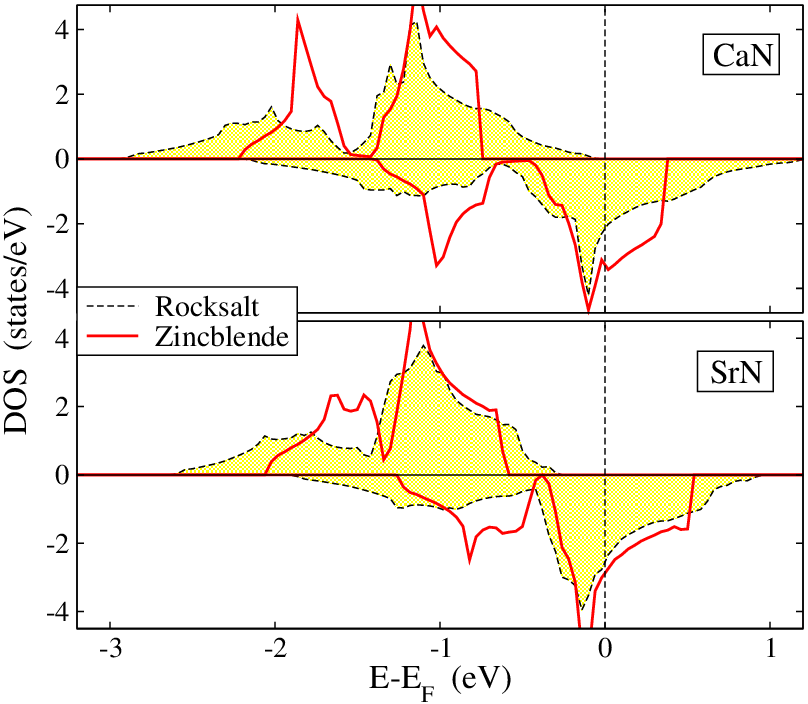}
\caption{(Color online) Total DOS of CaN and SrN alloys in both
the rocksalt (shaded region) and zincblende (red thick lines)
structures using the FPLO method.  Calculations are performed at
the calculated GGA equilibrium lattice constants (5.02 \AA\ and
5.45 \AA\ for CaN in the rocksalt and zincblende structures,
respectively, and 5.39 \AA\ and 5.82 \AA\ for SrN in the same two
lattice structures respectively). Details as in Fig. \ref{fig1}.}
\label{fig3}
\end{figure}

\begin{table*}
\caption{Calculated spin magnetic moments (in $\mu_B$) for SrN in
both rocksalt and zincblende structure using the FPLO method
within the GGA approximation. The  first line corresponds to the
ideal equilibrium lattice constants and the rest to
tetragonalization where we vary the in-plane lattice parameters by
the percentage shown in the second column and in the same time we
vary also the lattice parameter along the c-axis so that the
unit-cell volume is kept equal to the equilibrium one. Note that
in the second column the minus "-" sign means compression and the
plus "+" sign means expansion.  In all cases the total
spin-magnetic is kept equal to 1.0 $\mu_B$ and thus the
half-metallicity is preserved. Similar are the results for CaN,
where the only noticeable difference is that in all cases the
absolute values of Ca and N spin moments are smaller by about
0.01$\mu_B$ with respect to SrN and thus they sum up again to 1.0
$\mu_B$.} \label{table2}
\begin{ruledtabular}
 \begin{tabular}{lr|rrr|rrr}
&&\multicolumn{3}{c|}{SrN-RockSalt (5.39\AA ) }& \multicolumn{3}{c}{SrN-ZincBlende (5.82\AA ) }\\
     & Case & Sr & N & Total & Sr & N & Total\\ \hline
Ideal & & -0.074 &1.074 &1.000& -0.093 & 1.093 & 1.000  \\ \hline
Tetragonalization & -1\%  & -0.074&1.074 &1.000& -0.093 & 1.093 & 1.000 \\
& -5\%  & -0.073& 1.073&1.000 & -0.093 & 1.093 & 1.000\\
& -10\%  & -0.068& 1.068&1.000 & -0.093 & 1.093 & 1.000\\
&+1\%  &  -0.074&1.074 &1.000 & -0.093 & 1.093 & 1.000\\
&+5\% &  -0.072&1.072 &1.000  & -0.093 & 1.093 & 1.000\\
& +10\%  &  -0.069&1.069 &1.000& -0.094 & 1.094 & 1.000
\end{tabular}
\end{ruledtabular}
\end{table*}

\section{Stability of half-metallicity against lattice deformation}\label{sec4}

In the second part of our study we will focus on the stability of
half-metallicity against lattice deformations. To carry out these
calculations as stated in Section \ref{sec2} we have employed the
FPLO\cite{koepernik} code, which is  a full potential code and
thus is expected to describe tetragonal deformations more
accurately than the ASW\cite{asw} code which employs the atomic
sphere approximation.\cite{asa} Since we are interested in
deformations the first step is to calculate the equilibrium
lattice constants using total energy calculations. To this respect
we have employed the GGA approximation which is well-known to
produce more accurately results concerning the elastic properties
than the local-spin density approximation (LSDA). For CaN in the
rocksalt structure we found an equilibrium lattice constant of
5.02 \AA\ and for SrN 5.39 \AA . These values are almost identical
to the results of Geshi and collaborators who found for rocksalt
CaN and SrN 5.02 \AA\  and 5.37 \AA , respectively. To make our
study on deformations more complete we have also included the case
of zincblende CaN and SrN since the tetragonalization effect has
not yet been studied for this structure. Our FPLO-GGA total energy
calculations gave as an equilibrium lattice constant of 5.45 \AA\
for CaN and 5.82 \AA\ for SrN in the ZB-lattice. The lattice
constant of the ZB cubic unit cell is larger than in the
RS-structure since in the former case it also contains two void
sites while no voids are present in the RS-case. In Fig.
\ref{fig3} we present the total DOS for both compounds and for
both ZB and RS lattices at the equilibrium lattice constants. In
the RS-case our DOS within FPLO are similar to the ASW DOS in Fig.
\ref{fig1}.  In the ZB-cases the bands are more narrow than the
RS-structure and again both CaN and SrN are half-metallic
ferromagnets with a total spin magnetic in the formula unit of 1
$\mu_B$. For the ZB-lattice the half-metallic gaps are
considerable larger being around 0.5 eV for SrN and slightly
larger for CaN.

To simulate deformation of the lattices  with respect to
equilibrium we took into account the case of tetragonalization
where we vary
 both the in-plane and out-of-plane parameters keeping the unit
cell volume constant to the equilibrium. Such a deformation is
expected to model the growth of the materials under study on
different substrates. In Table \ref{table2} we have gathered our
results concerning the spin magnetic moments for SrN in both RS-
and ZB-lattices. We do not present the results for CaN since the
effect of deformation is identical for both alloys although their
electronic band structure in Fig. \ref{fig3} present slight
differences. In the second column we present the percentage of
change; "-" means compression by that percentage and "+" means
expansion, and the percentage refers to the in-plane lattice
constants. We took six values of the percentage into account:
$\pm$1 \%, $\pm$5 \% and $\pm$10 \%.. In all cases  the SrN
compound remains half-metallic in both the RS and ZB structures
with a total spin magnetic moment of 1 $\mu_B$. Deformations lead
to small changes of the absolute values of the Sr and N atomic
spin magnetic moments of less than 0.01 $\mu_B$ in such a way that
they cancel each other keeping the total spin moment constant.
Even in the case of CaN (not presented here) in the RS structure
where the half-metallic gap is only 0.1 eV (see discussion in
Section \ref{sec3}) the half metallic character is preserved even
for a $\pm$10 deformation. Our calculated DOS (note presented)
confirm the conclusion drawn from the spin magnetic moments and
are identical to the ideal cubic lattice for all degrees of
tetragonalization under study.

Finally we should also shortly comment on the expected behavior of
the Curie temperature upon tetragonalization.  In Refs
\onlinecite{Sasioglu05b} and \onlinecite{Kudrnovsky11} it was
shown by means of first-principles calculations  that the Curie
temperature in Heusler compounds increases with increasing
hydrostatic pressure which is in agreement with the initial
Castellitz interaction curve based on experiments for transition
metal compounds\cite{Castellitz} and its recent generalization for
Heusler alloys by Kanomata and collaborators.\cite{Kanomata} For
the sp-electron ferromagnets under study we expect a similar
behavior since in Ref. \onlinecite{Laref11} we have shown that the
Curie temperature is very sensitive to the lattice spacing in
zincblende pnictides and this also led to the larger Curie
temperature for CaN with respect to SrN (the former has smaller
lattice parameter). Tetragonalization keeping the volume constant
is not expected to change the Curie temperature. If, e.g.,  we
have smaller exchange parameters within the $xy$-plane with
respect to the equilibrium due to expansion, the compression in
the out of-plane $z$-axis will lead to larger exchange constants,
eventually compensating each other. Only in cases where strain
induces volume changes, like in the case of hydrostatic pressure
or tetragonalization keeping the in-plane lattice parameters
constant, one could expect variation in the estimated Curie
temperatures.

\section{Summary and conclusions}\label{sec6}

Half-metallic ferromagnets, which do not contain transition metal
atoms, are attractive for applications due to the smaller spin
magnetic moments. Among these so-called sp-electron ferromagnets
the case of alkaline-earth metal  mononitrides crystallizing in
the rocksalt, wurtzite or zincblende structures are promising
since the half-metallicity is combined with a total spin-magnetic
moment of 1 $\mu_B$ per formula unit leading to minimal energy
losses in spintronic applications. Moreover their equilibrium
lattice constant makes them suitable for growth on top of a
variety of semiconductors.

In the present study we have concentrated on the rocksalt
structure of CaN and SrN which is energetically favored with
respect to the wurtzite and zincblende ones. Employing
electronic-structure calculations in conjunction with the
generalized gradient approximation and using the frozen-magnon
technique we have studied the temperature dependent properties
upon varying the electron counting. Both alloys were found to
present Curie temperature above the room temperature (480 K for
CaN and 415 K for SrN using the random-phase approximation). Upon
small degrees of doping either with electrons or holes the Curie
temperature presented only a small decrease from its maximum
value, and we had to dope it with $\sim$0.6 electrons or $\sim$0.5
holes per formula unit for the Curie temperature to become
comparable to the room temperature. At the zero electron counting
(no doping) both the nearest and next-nearest N-N interaction
favored ferromagnetism; as we dope with electrons their intensity
decrease, while as we dope with holes the nearest N-N exchange
interaction becomes more sizeable while the next-nearest N-N
interaction starts favoring an antiferromagnetic configuration
leading to the decrease of the Curie temperature.

In the second part of our study we studied the response of both
the electronic and magnetic properties upon tetragonalization
keeping the unit cell volume constant. Except the rocksalt we
performed calculations also for the zincblende lattice for both
CaN and SrN. In all cases under study the half-metallicity was
preserved and both the electronic and magnetic properties only
marginally changed. Moreover, since the unit cell volume remains
constant, also the Curie temperature is not expected to vary.

Thus we can conclude that these materials are promising for
applications since small degrees of doping or large deformations
keep intact the half-metallic character and a high value for the
Curie temperature and their controlled experimental growth is
expected to boost the interest on sp-electron ferromagnets.


\begin{thebibliography}{99}


\bibitem{Zutic}
 I. \v{Z}uti\'c, J. Fabian, and S. Das Sarma, Rev.
 Mod. Phys. \textbf{76}, 323 (2004).

\bibitem{Felser}
C. Felser, G. H. Fecher, and B. Balke, Angew. Chem. Int. Ed.
\textbf{46}, 668 (2007).

\bibitem{Zabel}
H. Zabel, Materials Today \textbf{9},  42 (2006).

\bibitem{Katsnelson}
M. I. Katsnelson, V. Yu. Irkhin, L. Chioncel, A. I. Lichtenstein,
and R. A. de Groot, Rev. Mod. Phys. \textbf{80}, 315 (2008).


\bibitem{Volniaska10}
%Magnetism of solids resulting from spin polarization of p orbitals\\
O. Volnianska and P. Boguslawski, J. Phys.:Condens. Matter
\textbf{22}, 073202 (2010).

\bibitem{Zhou12}
%Possible room-temperature ferromagnetism in SnO2 nanocrystalline
%powders with nonmagnetic K doping\\
W. Zhou, X. Tang, P. Xing, W. Liu, and P. Wu, Phys. Lett. A
\textbf{376}, 203 (2012).


\bibitem{Uchino12}
T. Uchino and T. Yoko, Phys. Rev. B \textbf{85}, 012407 (2012).

\bibitem{Maca08}
%Magnetism without magnetic impurities in ZrO2 oxide\\
F. Maca, J. Kudrnovsk\'y, V. Drchal, and G. Bouzerar, Appl. Phys.
Lett. \textbf{92}, 212503 (2008).


\bibitem{Zheng11}
%Room-temperature ferromagnetism observed in alumina films\\
 Y. L. Zheng, C. M. Zhen, X. Q. Wang, L. Ma, X. L. Li, and D. L. Hou, Sol. St.
 Sci. \textbf{13}, 1516 (2011).

\bibitem{Guan10}
%Nonconventional magnetism in pristine and alkali doped In2O3:
%Density functional study \\
L. X. Guan, J. G. Tao, C. H. A. Huan, J. L. Kuo, and L. Wang, J.
Appl. Phys. \textbf{108}, 093911 (2010).

\bibitem{Ylvisaker10}
%Orbital order, stacking defects, and spin fluctuations in the
%p-electron molecular solid RbO2\\
E. R. Ylvisaker, R. R. P. Singh, and W. E. Pickett, Phys. Rev. B
\textbf{81}, 180405(R) (2010).

\bibitem{Kovacik09}
%Correlation effects in p-electron magnets: Electronic structure of
%RbO2 from first principles \\
R. Kovacik and C. Ederer, Phys. Rev. B \textbf{80}, 140411(R)
(2009).

\bibitem{Kim10}
%Antiferromagnetic and structural transitions in the superoxide KO2
%from first principles: A 2p-electron system with
%spin-orbital-lattice coupling\\
M. Kim, B. H. Kim, H. C. Choi, and B. I. Min, Phys. Rev. B
\textbf{81}, 100409(R) (2010).


\bibitem{Volniaska08}
%Molecular magnetism of monoclinic SrN: A
%first-principles study\\
O. Volnianska and P. Boguslawski, Phys. Rev. B \textbf{77},
220403(R) (2008).

\bibitem{Slipukhina11}
%Ferromagnetic Spin Coupling of 2p Impurities in Band Insulators
%Stabilized by an Intersite Coulomb Interaction: Nitrogen-Doped MgO\\
I. Slipukhina, Ph. Mavropoulos, S. Bl\"ugel, and M. Le\v{z}aic,
Phys. Rev. Lett. \textbf{107}, 137203 (2011).


\bibitem{Wu10}
%Magnetism in C- or N-doped MgO and ZnO: A Density-Functional Study
%of Impurity Pairs\\
H. Wu, A. Stroppa, S. Sakong, S. Picozzi, M. Scheffler, and P.
Kratzer, Phys. Rev. Lett. \textbf{105}, 267203 (2010).

\bibitem{Mavropoulos09}
%Ferromagnetism in nitrogen-doped MgO: Density-functional
%calculations\\
P. Mavropoulos, M. Le\v{z}aic, and S. Bl\"ugel, Phys. Rev. B
\textbf{80}, 184403 (2009).

\bibitem{Xiao10}
%Electronic structure and magnetic properties in Nitrogen-doped
%beta-Ga2O3 from density functional calculations\\
W.-Z. Xiao, L.-L. Wang, L. Xu, Q. Wan, and A.-L. Pan, Sol. St.
Commun. \textbf{150}, 852 (2010).

\bibitem{Banikov11}
%Magnetic and Electronic Properties of Nitrogen-Doped Lanthanum
%Sesquioxide La2O3 as Predicted from First Principles\\
V. V. Bannikov, I. R. Shein, and A. L. Ivanovskii, J. Supercond.
Nov. Magn. \textbf{24}, 1693 (2011).

\bibitem{Yang12}
%First-principles characterization of ferromagnetism in N-doped
%SrTiO3 and BaTiO3 \\
K. Yang, Y. Dai, and B. Huang, Appl. Phys. Lett. \textbf{100},
062409 (2012).


\bibitem{Kenmonchi04b}
%Materials Design of Transparent and Half-Metallic Ferromagnets of
%MgO, SrO and BaO without Magnetic Elements\\
K. Kenmochi, V. A. Dinh, K. Sato, A. Yanase, and H.
Katayama-Yoshida, J. Phys. Soc. Jpn \textbf{73}, 2952 (2004).

\bibitem{Kenmonchi04}
%New Class of Diluted Ferromagnetic Semiconductors based on CaO without Transition Metal Elements\\
K. Kenmochi, M. Seike,  K. Sato, A. Yanase, and H.
Katayama-Yoshida, Jpn. J. Appl. Phys. \textbf{43}, L934 (2004).

\bibitem{Liu11}
%Optical and magnetic properties of nitrogen ion implanted MgO
%single crystal\\
C. M. Liu, H.-Q. Gu, X. Xiang, Y. Zhang, Y. Jiang, M. Chen, and
X.-T. Zu Xiao-Tao, Chin. Phys. B \textbf{20}, 047505 (2011).

\bibitem{ReviewCrAs}
Ph. Mavropoulos and I. Galanakis, J. Phys. Condens. Matter
\textbf{19}, 315221 (2007).

\bibitem{Geshi04}
%Zinc-blende CaP, CaAs and CaSb as half-metals: A new route to
%magnetism in calcium compounds\\
M. Geshi, K. Kusakabe, H. Tsukamoto, and N. Suzuki, 2004 Preprint
arXiv:cond-mat/0402641 (2004).

\bibitem{Kusakabe04}
%New half-metallic materials with an alkaline earth element\\
K. Kusakabe, M. Geshi, H. Tsukamoto, and N. Suzuki, J.
Phys.:Condens. Matter \textbf{16}, S5639 (2004).


\bibitem{Liu08}
%Growth and self-assembly of MnN overlayers on Cu(0 0 1)\\
X. Liu, B. Lu, T. Iimori, K. Nakatsuji, and F. Komori, Surf. Sci.
\textbf{602}, 1844 (2008).


\bibitem{Sieberer06}
%Ferromagnetism in tetrahedrally coordinated compounds of I/II-V
%elements: Ab initio calculations\\
M. Sieberer, J. Redinger, S. Khmelevskyi, and P. Mohn,   Phys.
Rev. B \textbf{73}, 024404 (2006).

\bibitem{Zhang08}
%Half-metallic ferromagnetism in the zinc-blende MC (M = Li,
%Na and K)\\
C.-W. Zhang, J. Phys. D.: Appl. Phys. \textbf{41}, 085006 (2008).


\bibitem{Zberecki09}
%Ab initio prediction of half-metallic ferromagnetic metamaterials
%composed of alkali metals with nitrogen\\
K. Zberecki, L. Adamowicz, and M. Wierzbicki, Phys. St. Solidi (b)
\textbf{246}, 2270 (2009).

\bibitem{Yan12}
%Half-metallic properties in rocksalt and zinc-blende M N (M=Na,
%K): A first-principles study\\
E. Yan, Physica B \textbf{407}, 879 (2012).

\bibitem{Gao09}
%Bulk and surface sp half-metallic ferromagnetism in alkali metal
%pnictides with rocksalt structure: a first-principles
%calculation\\
G. Y. Gao, K. L. Yao, Z. L. Liu, Y. Min, J. Zhang, S. W. Fan, and
D. H. Zhang, J. Phys.:Condens. Matter \textbf{21}, 275502 (2009).


\bibitem{Gao11}
%Half-metallic ferromagnetism in rocksalt and zinc-blende MS (M=Li,
%Na and K): A first-principles study\\
 G. Y. Gao, K. L. Yao, M. H.
Song, and Z. L. Liu, J. Magn. Magn. Mater. \textbf{323}, 2652
(2011).


\bibitem{Gao07}
%Search for new half-metallic ferromagnets in zinc blende CaSi and
%CaGe by first-principles calculations\\
 G. Y. Gao, K. L. Yao, Z.
L. Liu, J. L. Jiang, L. H. Yu, and Y. L. Shi, J. Phys.:Condens.
Matter   \textbf{19}, 315222 (2007).

\bibitem{Gao07c}
%Half-metallic ferromagnetism in zinc-blende CaC, SrC, and BaC from
%first principles\\
 G. Y. Gao, K. L. Yao, E. \c{S}a\c{s}{\i}o\u{g}lu, L. M. Sandratskii, Z. L. Liu, and J. L.
Jiang, Phys. Rev. B \textbf{75}, 174442 (2007).

\bibitem{Gao07b}
%Half-metallic sp-electron ferromagnets in rocksalt structure: The
%case of SrC and BaC\\
G. Y. Gao and K. L. Yao, Appl. Phys. Lett. \textbf{91}, 082512
(2007).

\bibitem{Dong11}
%First-principles studies on magnetic properties of rocksalt
%structure MC (M = Ca, Sr, and Ba) under pressure\\
S. Dong and H. Zhao, Appl. Phys. Lett. \textbf{98}, 182501 (2011).

\bibitem{Zhang08b}
%Half-metallic ferromagnetism in wurtzite SrC\\
C.-W. Zhang, S.-S Yan, and H. Li, Phys. St. Solidi (b)
\textbf{245}, 201 (2008).

\bibitem{Gao09b}
%Surface sp half-metallicity of zinc-blende calcium monocarbide\\
G. Y. Gao  and K.-L. Yao, J. Appl. Phys. \textbf{106}, 053703
(2009).

\bibitem{Verma10}
%Ab initio studies of structural, electronic, magnetic and
%mechanical properties of alkali earth metal silicides\\
U. P. Verma, Mohini, P. S. Bisht, and P. Jensen, Semicond. Sci.
Technol. \textbf{25}, 105002 (2010).

\bibitem{Yao06}
%First principle prediction of half-metallic ferromagnetism in
%zinc-blende MBi (M= Ca, Sr, Ba) \\
K. L. Yao, J. L. Jiang, Z. L. Liu, and G. Y. Gao, Phys. Lett. A
\textbf{359}, 326 (2006).

\bibitem{Li08}
%Breakdown of half-metallic ferromagnetism in zinc-blende II-V
%compounds: First-principles calculations\\
Y. Li and J. Yu,  Phys. Rev. B \textbf{78}, 165203 (2008).


\bibitem{Volniaska07}
%Magnetic and structural properties of IIA-V nitrides\\
O. Volnianska and P. Boguslawski, Phys. Rev. B \textbf{75}, 224418
(2007).

\bibitem{Geshi07}
%Synthetic ferromagnetic nitrides: First-principles calculations of
%CaN and SrN\\
M. Geshi, K. Kusakabe, H. Nagara, and N. Suzuki, Phys. Rev. B
\textbf{76}, 054433 (2007).


\bibitem{Gao08}
%A first-principles study of half-metallic ferromagnetism in binary
%alkaline-earth nitrides with rock-salt structure\\
G. Y. Gao, K. L. Yao, Z. L. Liu, J. Zhang, Y. Min, and S. W. Fan,
Phys. Lett. A \textbf{372}, 1512 (2008).


\bibitem{Droghetti09}
%MgN: A possible material for spintronic applications\\
A. Droghetti, N. Baadji, and S. Sanvito, Phys. Rev. B \textbf{80},
235310 (2009).


\bibitem{Laref11}
%Exchange interactions, spin waves, and Curie temperature in
%zincblende half-metallic sp-electron ferromagnets: the case of CaZ
%(Z = N, P, As, Sb)\\
A. Laref, E. \c{S}a\c{s}{\i}o\u{g}lu, and I. Galanakis, J.
Phys.:Condens. Matter \textbf{23}, 296001 (2011).

\bibitem{Gao11b}
%Preserving the half-metallicity at the surfaces of rocksalt CaN
%and SrN and the interfaces of CaN/InN and SrN/GaP: a density
%functional study\\
G. Y. Gao, K. L. Yao, and N. Li, J. Phys.:Condens. Matter
\textbf{23}, 075501 (2011).


\bibitem{asw}
A. R. Williams, J. K\"ubler, and C. D. Gelatt, Phys. Rev. B
\textbf{19}, 6094 (1979).

\bibitem{asa}
O. K. Andersen, Phys. Rev. B \textbf{12}, 3060 (1975).

\bibitem{gga}
J. P. Perdew, K. Burke, and M. Ernzerhof, Phys. Rev. Lett.
\textbf{78}, 1396 (1997).

\bibitem{magnon}
L. M. Sandratskii and P. Bruno, Phys. Rev. B \textbf{66}, 134435
(2002).

\bibitem{Sasioglu}
E. \c{S}a\c{s}{\i}o\u{g}lu, I. Galanakis, L. M. Sandratskii, and
P. Bruno, J. Phys.:Condens. Matter \textbf{17}, 3915 (2005).

\bibitem{Galanakis}
I. Galanakis and E. \c{S}a\c{s}{\i}o\u{g}lu, J. Appl. Phys.
\textbf{109}, 113912 (2011).



\bibitem{koepernik}
K. Koepernik and H. Eschrig, Phys. Rev. B \textbf{59}, 1743
(1999).

\bibitem{Sasioglu05b}
E. \c{S}a\c{s}{\i}o\u{g}lu,  L. M. Sandratskii, and P. Bruno,
Phys. Rev. B \textbf{71}, 214412  (2005).

\bibitem{Kudrnovsky11}
S. K. Bose, J. Kudrnovsk\'y, V. Drchal, and I. Turek, Phys. Rev. B
\textbf{84}, 174422  (2011).

\bibitem{Castellitz}
L. Castelliz, A. Metallkde. \textbf{46}, 198 (1955).

\bibitem{Kanomata}
T. Kanomata, K. Shirakawa, and T. Kaneko, J. Magn. Magn. Mater.
\textbf{65}, 76 (1987).

\end{thebibliography}
\end{document}